# Hacia los Comités de Ética en Inteligencia Artificial

Sofía Trejo
Universidad Nacional Autónoma de México, 2019
sofia.trejo.a@gmail.com

Iván Meza
Instituto de Investigaciones en Matemáticas Aplicadas y en Sistemas,
Universidad Nacional Autónoma de México
ivanvladimir@turing.iimas.unam.mx

Fernanda López-Escobedo
Licenciatura en Ciencia Forense de la Facultad de Medicina
Universidad Nacional Autónoma de México
flopeze@unam.mx

**Resumen**

Los sistemas basados en inteligencia artificial tienen como finalidad la toma de decisiones que tienen efecto en el entorno. Ello apunta a que necesitamos mecanismos para regular el impacto que esta clase de sistemas tienen en la sociedad. Por esta razón, resulta primordial la creación de estatutos y organismos especializados que velen por el cumplimiento de los derechos humanos a nivel nacional e internacional. En este trabajo se propone la creación, en las universidades, de Comisiones y Comités de Ética en Inteligencia Artificial encargados de definir y garantizar el cumplimiento de principios de buenas prácticas en el campo.

**Introducción**

En la última década ha habido una revolución en el campo de Inteligencia Artificial que ha hecho posible transferir productos de investigación y desarrollos tecnológicos a diferentes campos de aplicación como el reconocimiento facial, la traducción automática y el reconocimiento de voz [1]–[3]. En particular, la digitalización de las comunicaciones y el alto acceso al internet han puesto estos avances literalmente en la palma de nuestras manos a través de los dispositivos móviles. Por ejemplo, hoy en día es muy común usar traductores automáticos en nuestro celular y con ellos descifrar mensajes en idiomas diferentes al nuestro. Hasta la década pasada, esta tecnología y los métodos asociados eran desarrollados incrementalmente en múltiples laboratorios de universidades alrededor del mundo. Con el advenimiento de las técnicas basadas en redes neuronales (también denominados *deep learning*) los ciclos de investigación, desarrollo, producción y adopción se han disminuido [4]. Por ejemplo, tomó a Google Translate un año portar las metodologías de redes neuronales que estaban siendo investigadas en 2015 para ponerlas a disposición de sus usuarios a mediados de 2016[1]. Aunque esto supondría buenas noticias para los ciudadanos digitales, la prontitud con que se hacen las investigaciones, conjugado con el alcance multitudinario que tienen estas tecnologías, apuntan a que necesitamos mecanismos formales basados en preceptos éticos para regular y planear el impacto que los sistemas basados en inteligencia artificial tienen en nuestra sociedad.

## Sistemas basados en Inteligencia Artificial

Los sistemas basados en inteligencia artificial son sistemas computacionales como lo son las hojas de cálculo, los editores de texto, las bases de datos, los sistemas de nóminas o las páginas de internet. Este tipo de sistemas, al igual que los ejemplos pasados, son ejecutados por computadoras y tienen una utilidad específica. A diferencia de otros sistemas computacionales, uno basado en inteligencia artificial tiene como fin la toma de decisiones y además, se espera que la decisión tomada tenga un efecto en el mundo real en diferentes niveles. Por ejemplo, existen sistemas que deciden comprar o vender acciones en el mercado de valores [5], la decisión a la que llega este sistema se espera que tenga un efecto directo promoviendo o descartando la compra de acciones, en algunas ocasiones miles de veces por segundo. En otros sistemas, la decisión tomada es para consumo de un usuario; por ejemplo, existen sistemas que deciden qué recomendar: productos, lecturas, películas, noticias, etc. [6]. Finalmente, existen sistemas que toman decisiones a bajo nivel como parte del proceso para producir una salida que será consumida por otro sistema. El reconocimiento de voz automático es de este tipo, las decisiones internas de estos sistemas producen una transcripción de lo dicho por el usuario, que después se usará como entrada por otro sistema [7].

Es deseable tener sistemas que tomen decisiones por nosotros dado su potencial de facilitarnos aspectos de nuestra vida cotidiana. Sin embargo, no hay que confundirlos con meras herramientas. Al caracterizarlos como herramientas negamos su propiedad más definitoria que es su capacidad de actuar en el mundo, es decir, son sistemas con un nivel de agencia. La razón de ser de un sistema basado en inteligencia artificial es tomar decisiones en situaciones "reales". Es cuando este ciclo se cierra que los sistemas son de mayor beneficio para nosotros, por ejemplo, el sistema que detecta si un correo electrónico es de tipo *spam* o no, es útil hasta que aplicamos ese sistema a los correos que llegan a nuestra bandeja; el efecto es tener orden

---

[1] El artículo *The great AI Awakening* narra el ciclo de desarrollo para producir la primera generación de traductores automáticos basados en redes neuronales por la empresa Google:
https://www.nytimes.com/2016/12/14/magazine/the-great-ai-awakening.html (visitado 28 de noviembre de 2019)

en nuestros correos y por ende en nuestras vidas digitales. Sin embargo, es su agencialidad la que hace que tengamos que poner mayor atención al impacto que tiene el uso masivo de estos sistemas.

**Recientes avances en la Inteligencia Artificial**

La razón principal del avance en el campo de la inteligencia artificial en la última década se debe a la adopción y explotación de los sistemas de redes neuronales, ahora llamados de *aprendizaje profundo [8]*. Esta adopción ha sido motivada por un cambio fundamental en el pensamiento de cómo construir sistemas inteligentes y de la promesa de una mejora económica para los individuos u organizaciones que desarrollen este tipo de sistemas. En particular, en la última década, el campo se ha enfocado en la construcción de sistemas que imitan comportamientos inteligentes en lugar de crear sistemas que posean conocimiento y estrategias que guíen el comportamiento inteligente. Esta preferencia por sistemas imitadores de inteligencia ha sido gradual y se sustenta en tres factores fundamentales:

1. *La consolidación del campo de aprendizaje de máquina:* el aprendizaje de máquina es un campo de estudio de modelos que capturan comportamientos codificados en ejemplos de ese comportamiento. Por ejemplo, para decidir si un correo corresponde a *spam* o no, se necesitan cientos o miles de correos con la especificación de si es o no *spam.* Los métodos de aprendizaje máquina pueden identificar patrones que diferencian correos *spam* de no spam. En particular, el aprendizaje profundo es un tipo de aprendizaje de máquina.
2. *La disponibilidad masiva de datos:* el uso de metodologías de aprendizaje profundo impone que para crear un sistema inteligente que imita un comportamiento es necesario contar con ejemplos de este comportamiento. En conjunto con la consolidación del aprendizaje automático vino el advenimiento de la circulación y almacenamiento de datos en forma masiva, motivada por internet, avances en los campos de *big data* y telecomunicaciones.
3. El incremento de la capacidad de cómputo: en particular, el procesamiento de las cantidades masivas de datos por las técnicas de aprendizaje automático requieren una capacidad de cómputo considerable. Afortunadamente, la capacidad de cómputo ha crecido siguiendo la ley de Moore [9], no sólo eso sino que en los últimos años ha surgido equipo específico de *hardware* que apoya la aceleración de los métodos de aprendizaje profundo[2].

---

[2] Las Unidades de Procesamiento Gráfico (GPUs) han sido adoptadas para su uso en la creación de sistemas inteligentes. También han surgido soluciones específicas como TPUs y arquitecturas FPGAS.

Estos tres factores han permitido una revolución en el campo de la inteligencia artificial, organizaciones que tienen acceso al equipo de cómputo adecuado, que dominan la técnica de aprendizaje profundo y poseen los datos adecuados están en la posición de crear sistemas que imitan la inteligencia: traductores, reconocedores de voz, *traders* automático, etc. Se ha preferido este enfoque sobre la creación de sistemas con inteligencia dado que para muchos comportamientos inteligentes no se cuenta con una teoría o con un conocimiento de partes de ésta. Ante dicha incertidumbre, la creación de sistemas inteligentes usando esta metodología de aprendizaje automático ha permitido llevar a usuarios finales estas capacidades.

Sin embargo, el avance en inteligencia artificial viene con retos asociados a los factores que han motivado al campo. Un reto es económico, tener acceso al equipo de cómputo ideal para desarrollar buenos sistemas es fundamental pero no es posible para todos, ya que los equipos no son accesibles. Otro reto tiene que ver con el dominio de la técnica, grupos de investigación y organizaciones que fueron rápidos en adoptar y dominar las técnicas hoy conforman centros de vanguardia en el campo, para el resto, hoy en día es una carrera de actualización continua. Finalmente, otro reto es la adquisición de datos. De hecho, este es el reto más estratégico ya que como veremos no es sólo el hecho de contar con los datos sino las propiedades intrínsecas de éstos las que determinan el tipo de sistema a crear, y hoy en día las mayores deficiencias de los sistemas inteligentes modernos recaen en que no tienen acceso a los datos adecuados o simplemente no se identificó a los datos como elemento esencial.

## Problemas con la naturaleza de datos

Dado que la metodología dominante requiere datos que contengan el comportamiento a imitar, el proceso asociado a conseguir esos datos es esencial al desempeño de los sistemas. La recolección, limpieza y preprocesamiento de los datos que alimentan a los sistemas inteligentes son una de las tareas que consumen más tiempo en la etapa de desarrollo. Es interesante que conceptos tan establecidos como "garbage in garbage out" tienen un efecto palpable en el campo de sistemas inteligentes. Si no cuidamos la recolección, la limpieza y el preprocesamiento de los datos, nuestro sistema tomará decisiones sin sentido, poco útiles y no benéficas para nosotros. Adicional a esta complejidad técnica se presenta un problema más sutil que son los sesgos ocultos de los datos y que son replicados por el comportamiento del sistema inteligente [10]. Estos sesgos muchas veces no tienen que ver con la calidad de los datos sino con estructuras sociales que permean hacía los datos.

Una de los casos más conocidos son los sesgos en el nivel de desempeño de los sistemas de reconocimiento de rostros automáticos. Un sistema automático de reconocimiento de rostros tiene el objetivo de identificar de forma automática propiedades sobre un rostro que observa, algunas de esas propiedades podrían ser: edad o género. En su trabajo *Buolamwini and Gebru* identifican el sesgo sistemático que tienen reconocedores de rostro comerciales al tratar de identificar el género de un rostro[3]: mientras que hombres con piel blanca son identificados en un 100%, mujeres con piel negra son clasificadas de forma correcta entre un 68.7% o un 77.5% [11]. Esta disparidad en la certeza sólo se justifica por el tipo diferente de dato de entrada, hombre blanco vs mujer de color; sin embargo, dado que el algoritmo y sistema es el mismo, la disparidad sólo puede tener origen en los datos con los que se *entrenó*. En esos datos se ha encontrado una subrepresentación de mujeres de color, ante este sesgo el sistema "aprende" a replicar este sesgo en su comportamiento.

Combatir los sesgos de género se ha convertido en una prioridad del campo de la inteligencia artificial, sin embargo como están embebidos en los comportamientos propios de nuestra sociedad, los esfuerzos han sido fútiles. Recientemente circuló en las noticias tecnológicas la decisión de una compañía internacional de parar el desarrollo de un sistema que apoya la recomendación de candidatos a un puesto basado en el currículum[4]. En particular se buscaba aprovechar toda la experiencia disponible en base de datos gigantescas de CVs y los casos de éxito de contratación; sin embargo, un poco de reflexión sobre las primeras versiones de este sistema hizo evidente que este sistema "aprendía" a favorecer curriculums con lenguaje típico de candidatos hombres y discriminación a mujeres. Otra vez, el sesgo presente debido a décadas de experiencia contratando a candidatos hombres sobre mujeres se hacía presente en el comportamiento "aprendido" por el sistema inteligente.

Como vemos, por un lado las relaciones intrínsecas de los datos reflejan muy bien estructuras no explícitas de la sociedad, como lo es la discriminación; y por otro lado, nuestra técnica principal aprende a imitar el comportamiento, el sistema resultante imita los comportamientos discriminatorios.

## Problemas con las fuentes de datos

Es evidente que los datos son esenciales para el desarrollo tecnológico de sistemas inteligentes, y aunque hay una gran disponibilidad de datos actualmente, no todos las grandes colecciones de datos son igualmente útiles. En particular, las colecciones que sobresalen son aquellas en donde las entradas están etiquetadas con su salida correspondiente. Estas son de mucho valor ya que nuestras mejores técnicas son particularmente buenas en este esquema (llamado *supervisado*). En particular, bajo este esquema es posible crear un sistema que imite el comportamiento codificado en la entrada-salida. Y aunque otros esquemas donde no se

---

[3] Información adicional sobre la metodología y actualizaciones de los experimentos se pueden encontrar en el siguiente sitio http://gendershades.org/ (visitado 28 de noviembre de 2019)

[4] Nota original sobre la suspensión del desarrollo del sistema: https://www.reuters.com/article/us-amazon-com-jobs-automation-insight/amazon-scraps-secret-ai-recruiting-tool-that-showed-bias-against-women-idUSKCN1MK08G (visitado 28 de noviembre de 2019)

necesita etiquetar la salida han comenzado a ser muy exitosos en el último lustro (aprendizaje por refuerzo [12] o *self-supervised [13]*) los sistemas supervisados son los que han permitido una nueva generación de sistemas traductores, reconocedores de voz, reconocedores de rostros, etc.

Algunas preguntas válidas bajo el esquema de supervisión son *¿de dónde salen estas colecciones? ¿qué procedimientos se siguieron para su recolección? ¿qué garantías proporcionan estas colecciones?* Para poder comenzar a responder estas preguntas tenemos que entender que el valor asociado a estas colecciones viene de dos lados: de la secrecía de los datos, estratégicamente mantenerlos secretos da valor al sistema inteligente que imita el comportamiento; por otro lado, apertura de la colección también le da valor, ya que al poder ser utilizada por muchos grupos de desarrollo permite incrementalmente ir aumentando la capacidad de imitación de los sistemas. Ante esta dicotomía hay colecciones de datos que son valiosas porque son secretas y hay colecciones que son valiosas porque son abiertas. En el caso de las colecciones abiertas surgen de la comunidad científica que las hacen disponibles para avanzar el campo, generalmente se siguen procedimientos que garantizan una cobertura amplia pero las garantías de éstas son pobres, lo más que se busca es que sean en un contexto definitivo representativas de una población. Como vimos en la sección anterior el hecho de que sean representativas también garantiza que vengan acompañada de sesgos.

Igualmente preocupante a los sesgos presentes en los datos es que en el afán de recolectar las colecciones de datos, algunas preceptos éticos no son tomados en cuenta. En los últimos años ha salido a la luz que las bases de datos se recolectaron sin consentimiento informado de las personas tomando en cuenta legislaciones o contextos legales laxos. La comunidad de visión computacional, encargada de hacer sistemas inteligentes con aplicaciones como la identificación de personas, el reconocimiento de éstas y su monitoreo, ha estado plagado de errores. La investigación MegaPixels, comandada por Harvey and LaPlace, expone algunos de estos errores, visualiza el impacto de estas colecciones e invita a la reflexión sobre éstas [14]. Hasta ahora tres colecciones masivas de imágenes han retirado su acceso público. Este año la organización detrás de ImageNet, una de las colecciones de imágenes más amplias, ha estado trabajando en arreglar aspectos de justicia y representatividad en la colección [15]

En el caso de las colecciones secretas es más difícil tener acceso e identificar omisiones éticas dada su naturaleza privada. Sin embargo, algunos casos han sido puestos a la luz. Recientemente se puso a la luz que una empresa desarrolló una aplicación móvil para guardar y organizar fotos personales en la nube, después la misma empresa usó fotos de sus usuarios para entrenar una herramienta de reconocimiento de rostros que forma parte de otro producto que ofrece con el propósito de monitoreo[5], lo anterior bajo el sustento que los *términos de uso* de la primera aplicación lo permiten. Uno de los casos más notables sobre la extracción de información sin consentimiento fue a través de la plataforma Facebook que permitió a asociados terceros recolectar una cantidad gigantesca de información que luego fue usada con fines electorales [16]. México no quedó exento de estas prácticas de recolección masiva de datos[6]

**Problemas en el propósito de los sistemas**

La metodología de creación de sistemas inteligentes basados en técnicas de aprendizaje de máquina no impone ninguna restricción en qué es la entrada y qué es la salida esperada. Uno puede poner cualquier lenguaje como entrada y como salida cualquier otro, y con suficientes ejemplos y un poco de adaptación el sistema podrá aprender a traducir de una lengua a otra. Igualmente, uno puede destinar este mecanismo a cualquier tarea. La tentación de poner cualquier elemento es muy alta, recientemente investigadores usaron fotografías de presos contra fotografías de personas en redes sociales para determinar factores criminalísticos en los rostros, por supuesto, estas ideas se encontraron con múltiples críticas éticas[7]. Otros casos no ha sido así y han terminado en investigaciones publicadas: autogenerar comentarios en redes sociales [17] y decidir el veredicto de juicios con base a hechos del expediente [18].

Para sistemas comerciales el campo de sistemas inteligentes abre un abanico de posibilidades, en particular estamos ante la oportunidad de modelar cualquier comportamiento, en particular sistemas comerciales están interesados en aquellos comportamientos que no sólo les dan una ventaja competitiva sino que hacen que sus usuarios no abandonen las plataformas. Por ejemplo, no es secreto que empresas buscan mejores recomendaciones ya que el usuario disfrutará el servicio y permanecerá más tiempo en la plataforma. Hoy en día se investigan mecanismos de aprendizaje automático que repercuten en tiempo viendo contenido en las plataformas [19]. Sin embargo, si la plataforma siempre es capaz de proporcionar contenido interesante al usuario, para éste es muy difícil abandonar la plataforma. Actualmente, se habla de mecanismos de adicción asociados a los comportamientos de consumo de estos servicios [20].

---

[5] Nota sobre uso de fotos personales para entrenar sistema de reconocimiento de rostros https://www.engadget.com/2019/05/09/ever-photo-storage-app-facial-recognition/ (visitado 30 de noviembre de 2019).

[6] Fuga de datos de empresa de medios en México https://www.upguard.com/breaches/facebook-user-data-leak (vistado 230 de noviembre de 2019).

[7] Crítica a sistema basado en imágenes de rostros para determinar criminalidad https://callingbullshit.org/case_studies/case_study_criminal_machine_learning.html (visitado 30 de noviembre de 2019)

## Inteligencia Artificial y Derechos Humanos

Los problemas sociales asociados con las inteligencias artificiales actuales tienen dos fuentes principales: los datos y el diseño. Como vimos anteriormente, los datos utilizados durante la creación de sistemas autónomos son considerados por los mismos como "la realidad". Esto quiere decir que, de haber sesgos en los datos, éstos son repetidos y amplificados por el sistema, lo cual está teniendo como consecuencia la sistematización de problemáticas sociales como el racismo y la discriminación de género [21]. Esto evidencia que los sistemas autónomos atentan contra los derechos de igualdad y no discriminación. En cuanto al diseño, la mayoría de sistemas inteligentes tienen como objetivo optimizar una función específica. Esto se convierte en un problema cuando el objetivo para el cual fue diseñado un sistema no es considerado benéfico para la sociedad. Para entender mejor esta problemática pensemos en una inteligencia artificial encargada de decidir qué contenido es mostrado a los usuarios en las redes sociales. En otras palabras, el sistema tiene como objetivo la moderación de contenido. El diseño más utilizado para este tipo de sistemas es el de optimizar el número de clics [22]. Esto quiere decir que el sistema da prioridad al contenido que ha sido comentado o compartido muchas veces. Ahora bien, gran parte del contenido viral en internet tiende a ser polémico, extremista y a generar un discurso de odio. De este modo, crear sistemas autónomos con la función de optimizar clics da prioridad a cierto tipo de discursos sobre otros [23]. Lo que nos obliga a hacer una seria reflexión sobre la influencia que la inteligencia artificial tiene sobre la libertad de expresión y la libre autodeterminación de las personas. Estas problemáticas son sólo algunos ejemplos del por qué necesitamos establecer, de manera urgente, normativas y principios de buenas prácticas en el área de inteligencia artificial, que permitan garantizar la protección de los derechos humanos.

## Propuesta para la creación de Comisiones y Comités de Ética en Inteligencia Artificial

La creación de estatutos y organismos especializadas en la regulación y el arbitraje en el campo de la inteligencia artificial a nivel nacional e internacional será un proceso complejo, el cual requerirá la colaboración del gobierno, la industria, la academia y la sociedad civil. Sin embargo, tomando en consideración que gran parte de las técnicas y los avances en el área son llevados a cabo en universidades, es esencial que dichas instituciones comiencen a regular los sistemas inteligentes creados dentro de sus instalaciones. Para ello, proponemos que cada universidad nacional, donde se desarrolle inteligencia artificial, integre una comisión y un comité de ética, ambos especializados en las problemáticas asociadas al desarrollo de dicha tecnología. Consideramos que la función primordial de las *Comisiones de Ética en Inteligencia Artificial* debe ser la de definir los principios de buenas prácticas para el desarrollo de sistemas inteligentes dentro de cada institución. De manera ideal, las *Comisiones de Ética en Inteligencia Artificial* de las universidades estarán en comunicación y colaboración constante, para homogeneizar la regulación del campo a nivel nacional. Por otro lado, sugerimos que el rol de los *Comités de Ética en Inteligencia Artificial* sea el de vigilar el cumplimiento de los principios éticos establecidos en la institución correspondiente, a través de la evaluación y el seguimiento de proyectos relacionados con inteligencia artificial.

Debido a las capacidades y la amplia variedad de aplicaones de sistemas autónomos, creemos esencial que las comisiones y comités de ética en inteligencia artificial sean integradas por especialistas en distintas áreas del conocimiento. En particular, es fundamental crear vínculos estrechos entre los organismos que regulen la inteligencia artificial y los comités de bioética existentes a nivel nacional. De hecho, creemos que establecer el nexo entre la bioética y la ética en inteligencia artificial es el paso central del proceso de estandarización de buenas prácticas en el área. Sin embargo, es importante recalcar que existen diferencias significativas entre las problemáticas dentro del ámbito de la bioética, y las relacionadas con la inteligencia artificial. Para entender la relación entre estas dos áreas, y cómo es que la bioética juega un papel central en nuestra propuesta debemos discutir el tema de la bioética más a fondo.

## La bioética como referente

La velocidad vertiginosa con la que progresan la ciencia y la tecnología genera un constante desfasamiento entre el uso de ciertas técnicas y su regulación. Un claro ejemplo de esto se puede apreciar en el campo de la biotecnología. Esta área, dedicada al uso de seres vivos y organismos para la creación de productos ha existido desde tiempos inmemoriales. La selección artificial de plantas que produjo la agricultura, la fermentación y la producción de vacunas son ejemplos de avances en el campo. En décadas más recientes, la biotecnología ha desarrollado procedimientos para la edición de ADN y la síntesis de proteínas. La capacidad de manipular los bloques fundamentales de la vida expandió nuestros horizontes, dándonos la posibilidad de recuperar especies extintas o de crear nuevos medicamentos. Sin embargo, también amplificó la capacidad humana de producir impactos negativos en el planeta. Por ejemplo, modificar genéticamente a los mosquitos para erradicar enfermedades como la malaria podría tener efectos catastróficos en el medio ambiente [24]. Este tipo de problemáticas, de igual modo las derivadas de la biología y de las ciencias de la salud, han llevado a la creación de normativas y principios de buenas prácticas que tienen como eje central a la bioética. Esta disciplina se encarga de ampliar las consideraciones morales, tradicionalmente asociadas a los humanos, a los demás seres vivos. En otras palabras, la bioética tiene como interés central el bienestar no sólo de la especie humana, sino de todos los organismos que habitan en el planeta. Actualmente, existen organismos nacionales (*e.g.* CONBIOÉTICA) e internacionales (*e.g.* UNESCO) enfocados en el desarrollo e implementación de la bioética, asimismo, las universidades cuentan con comités de bioética que regulan las actividades clínicas y académicas.

Para dar un ejemplo de las normativas bioéticas existentes a nivel universitario, tomaremos como referente a la Universidad Nacional Autónoma de México. Usando como fuente el acuerdo publicado en agosto del 2019 donde es establecen los lineamientos para la integración y conformación de comités de ética en la universidad[8]. Este documento define las responsabilidades de los comités de bioética de la siguiente manera:

**Comité de Bioética:** *se encarga de abordar sistemáticamente y de forma continua la dimensión ética de a) las ciencias médicas y de la salud, b) las ciencias biológicas y c) las políticas de salud*[9].

En particular, los lineamientos referentes a investigaciones y prácticas donde participan seres humanos establecen que dichas actividades deben responder a los siguientes principios éticos[10]:

1. **Principio de autonomía:** *implica el reconocimiento de la capacidad del sujeto de la investigación para la toma de decisiones que se plasma en el consentimiento informado.*
2. **Principio de no maleficencia:** *se entiende como la responsabilidad del personal académico o alumnado de minimizar los daños y riesgos reales o potenciales de quienes participan como sujetos de investigación.*
3. **Principio de beneficencia:** *implica maximizar los beneficios para los sujetos participantes en la investigación, además de garantizar que el riesgo de las investigaciones sólo puede tomarse cuando no exista una alternativa.*

Creemos que los principios éticos de autonomía, no maleficencia y beneficencia, centrales para la bioética, deben ser adoptados dentro del campo de la inteligencia artificial. En particular, es fundamental que las investigaciones y los proyectos de inteligencia artificial cuenten con el consenso informado de los individuos de quienes fueron recolectados los datos. Normativa que hasta ahora no existe en la mayoría de los campos de la ciencia de datos.

## Recomendaciones

Existen sistemas de inteligencia artificial que entran dentro de las regulaciones de la bioética, como los que utilizan datos de voz. Sin embargo, la ética de la inteligencia artificial debe atender problemas sociales, como la discriminación y la libertad de expresión, dando prioridad al respeto de los derechos humanos. Por ello, consideramos fundamental que se reflexione y trabaje para crear una ética de la inteligencia artificial, nutrida de la bioética, pero enfocada al estudio de las problemáticas derivadas de la ciencia de datos. Crucial para esto será la consolidación de las comisiones y los comités de ética universitarios.

---

[8] Gaceta UNAM, 29 de agosto 2019, https://www.gaceta.unam.mx/wp-content/uploads/2019/10/190829-Convocatorias.pdf (visitado 6 de diciembre de 2019).

[9] Tomado de *idem 8.*

[10] Tomados de *idem* 8.

A continuación hacemos un análisis de la legislación universitaria que consideramos relevante para la creación de *Comités de Ética de la Inteligencia Artificial* en las universidades nacionales. En particular, utilizaremos como referente el acuerdo publicado en la Gaceta UNAM en agosto del 2019 que establece los lineamientos para la conformación y registro de comités de ética en la Universidad Nacional Autónoma de México. Documento donde se establece que los Comités de Ética deben ser órganos especializados con el compromiso de vigilar las prácticas académicas y científicas garantizando el respeto y la protección de los sujetos de investigación, procurando que se actúe de acuerdo a las buenas prácticas, y que atiendan las cuestiones éticas de su respectiva dependencia universitaria. Dentro de esta definición se propone hacer una distinción de Comités de Ética de acuerdo con las tareas específicas que sean llevadas a cabo por el organismo. Existiendo comités con las funciones: consultiva, de dictamen de seguimiento y educativa. En particular, de acuerdo con nuestra propuesta, los *Comités de Ética en Inteligencia Artificial* deberían tener la función de dictamen, ya que dentro de los lineamientos universitarios los comités con la función de dictamen son los encargados de analizar, revisar y dictaminar desde la perspectiva ética los protocolos de investigación de disciplinas donde se incluya la participación de seres humanos. Por otro lado, cabe destacar que dentro de los acuerdos para la creación de comités de ética de la UNAM no se proponen la creación de normativas y de principios de buenas prácticas como una de las funciones a ser desempeñadas por los Comités de Ética. Es por ello que hemos propuesto la creación de un organismo distinto encargado de desempeñar dicha función: las Comisiones de Ética en Inteligencia Artificial. Cabe destacar, que la universidad requiere que los Comités de Ética redacten una Guía de Funcionamiento Interno, sin embargo este documento hace referencia a la parte operacional del comité, más que atender a los cuestionamientos éticos de su área de especialización.

Continuando con el análisis del acuerdo universitario existen distintos tipos de comités, clasificados de acuerdo al tipo de problemáticas a abordar. Por ejemplo, existen comités de Bioética, de Bioseguridad, de Integridad Académica y Científica, entre otros. En particular, nosotros sugerimos que el *Comité de Ética en Inteligencia Artificial* sea considerado como un Comité de Ética de la Investigación, definido como[11]:

> **Comité de Ética de la Investigación:** *Es un tipo de comité que se conforma en las coordinaciones, entidades académicas y dependencias universitarias en donde se llevan a cabo actividades de investigación o docencia en seres humanos, los cuales participan como sujetos de investigación, en el marco de las disciplinas de las ciencias exactas, naturales, biológicas, humanidades y sociales, con el objetivo de salvaguardar sus derechos.*

Creemos que este tipo de comité ético es el más adecuado para representar el campo de la inteligencia artificial, ya que consideramos que gran parte de los sistemas autónomos actuales caen dentro de la definición de investigación en seres humanos. La cual está definida de la siguiente manera[12]:

---

[11] Tomado de *idem* 8.
[12] Tomado de *idem* 8.

> ***Investigación en y/o con seres humanos:*** *involucra la recopilación o análisis sistemático de información sobre seres humanos, con el propósito de generar nuevos conocimientos en la que los sujetos de investigación son expuestos a la manipulación, intervención, observación u otra interacción con personal académico, ya sea de manera directa o a través de la alteración de su entorno.*

Sabemos que existen sistemas autónomos que recopilan y analizan información sobre usuarios de manera constante. En particular, los motores de búsqueda, los sistemas de recomendación de películas y los traductores en línea son ejemplos de ello. Además de analizar información, estos sistemas toman decisiones y actúan en el mundo, muchos de ellos en tiempo real. Por ejemplo, la inteligencia artificial modera el contenido de las redes sociales utilizando los *clics* de los usuarios para seleccionar el contenido que será presentado en las plataformas. Esto quiere decir que usan los datos obtenidos de humanos para alterar su entorno. Por ende, existen varios sistemas autónomos que deben ser clasificados como investigación en seres humanos. Las normativas asociadas a este tipo de procedimiento deben ser respetadas para el desarrollo de sistemas de inteligencia artificial. En particular, se deberían respetar los principios éticos de autonomía, no maleficencia y beneficencia. Lo cual incluye la obligación de contar con un consentimiento informado por parte de los individuos cuyos datos son utilizados para la creación de sistemas inteligentes. Por otro lado, el principio de no maleficencia debería garantizar que los sistemas de inteligencia artificial sean incluyentes y que actúen de manera equitativa dentro de distintas poblaciones. Ya que, como hemos mencionado, existen sistemas autónomos que discriminan a grupos de personas.

Es importante recalcar que varias de las problemáticas derivadas de los sistemas autónomos están relacionadas con los datos que son generados por los humanos, sin embargo no queda claro cuáles son las limitantes para que esta información se considere como parte de la investigación en seres humanos. Determinar qué datos pueden ser utilizados libremente, cuales requieren de un consentimiento informado, qué información debería ser considerada privada o sensible será parte de las tareas de las *Comisiones de Ética en Inteligencia Artificial*. En conclusión, debemos encontrar la forma de traducir los dilemas éticos asociados al desarrollo de inteligencia artificial al lenguaje propio del campo, para que se controlen las problemáticas con los datos y el diseño de dichos sistemas. Para ello debemos de re-interpretar los principios de la bioética en el contexto de la inteligencia artificial para salvaguardar los derechos humanos, ya que esta es la única forma de garantizar que dichos sistemas actuen para el beneficio de la sociedad.

### Referencias: